\documentclass[11pt]{article}
\usepackage{moriond,epsfig}

\def\lsim{\mathrel{\rlap{\lower4pt\hbox{\hskip1pt$\sim$}}
     \raise1pt\hbox{$<$}}}         
\def\gsim{\mathrel{\rlap{\lower4pt\hbox{\hskip1pt$\sim$}}
     \raise1pt\hbox{$>$}}}         

     \def \thl {{\theta_l}}
\def \thK {{\theta_{K^*}}}

\def\be{\begin{equation}}
\def\ee{\end{equation}}
\def\bea{\begin{eqnarray}}
\def\eea{\end{eqnarray}}

\begin{document}
\title{
THE PHENO-ANALYSIS  OF $B \to K^{(*)}\mu^+ \mu^-$ DECAYS IN 2011 PLUS}
DO-TH 11/14

\author{G.HILLER}

\address{
Institut f\"ur
  Physik, Technische Universit\"at Dortmund, D-44221 Dortmund, Germany}
  
\maketitle\abstracts{
We report on  recent developments in the phenomenology of exclusive 
$b \to  s \mu^+ \mu^-$ decays in
testing the standard model and explore its borders: the benefits  of 
the region of large dimuon invariant masses and the exploitation of the angular distributions.
Consequences of model-independent analyses from current and future data are pointed out.}

\section{Introduction}

There exists a strong and long-standing interest  in 
$\Delta B=1$ exclusive $b \to s \mu^+ \mu^-$ processes  because
of their accessibility at hadron colliders, good theory control and sensitivity to
short-distance physics with and beyond the standard model, see, for instance \cite{Ali:1999mm}
\cite{Bobeth:2009ku}. Many modes have been observed by now by several experiments with
branching ratios at the level of  $10^{-(6-7)}$, such as
$B \to K^{(*)} \mu^+ \mu^-$   by BaBar \cite{Aubert:2008ps}, Belle \cite{:2009zv}, CDF \cite{Aaltonen:2011cn},  and recently
$B_s \to \Phi \mu^+ \mu^-$  decays by   CDF  \cite{Aaltonen:2011cn}.

At present, each experiment has collected about order hundred events per mode.
This already enables dedicated studies of spectra and asymmetries \cite{Bobeth:2010wg} \cite{Bobeth:2011gi}, which exhibit  a much larger sensitivity to electroweak physics than the determination of the (un-binned integrated) branching ratios.
The situation will further improve in the near future with the anticipated updates from the $b$-factories and the Tevatron, and with the ongoing run of the LHC.
In fact, LHCb has reported  35 
$B^+ \to K^+ \mu^+ \mu^-$ events in 37 pb$^{-1}$, and expects by the end of 2011 with 1 fb$^{-1}$  order $10^3$ $B \to K^* \mu^+ \mu^-$ events \cite{GolutvinLaThuile}.

\section{$B \to K^{*}\mu^+ \mu^-$ Theory and Recent Highlights }

The kinematically available phase space in $B \to K^{(*)}\mu^+ \mu^-$  decays
is $4 m_\mu^2 \leq q^2 < (m_B -m_{K^{(*)}})^2$ for the dilepton invariant mass squared $q^2$.
This region is fully covered experimentally with the exception of the 
$J/\Psi$ and $\Psi^\prime$ resonances  from
$B \to K^{(*)}  (\bar c c ) \to K^{(*)}  \mu^+ \mu^-$, which are removed by cuts.
A systematic theory treatment exists for the region of high $q^2 \sim {\cal{O}}(m_b^2)$
 by means of an operator product expansion (OPE)
put forward by Grinstein and Pirjol some time ago \cite{Grinstein:2004vb}.
The latter approach has recently been phenomenologically developed and exploited \cite{Bobeth:2010wg} \cite{Bobeth:2011gi}.
We give a brief  overview of the benefits of the high-$q^2$ region, corresponding to low hadronic recoil, in Section \ref{subsec:loreco}.
In the region of  low $q^2$, where the $K^{(*)}$ is energetic
in the $B$ restframe, the decays are eligible to QCD factorization methods 
\cite{Beneke:2004dp}. The region between the $J/\Psi$ and the $\Psi^\prime$ peaks
is informative on charmonia physics \cite{KMPW}.

\begin{figure}
\begin{center}
\vskip -0.0truein
{\epsfig{file=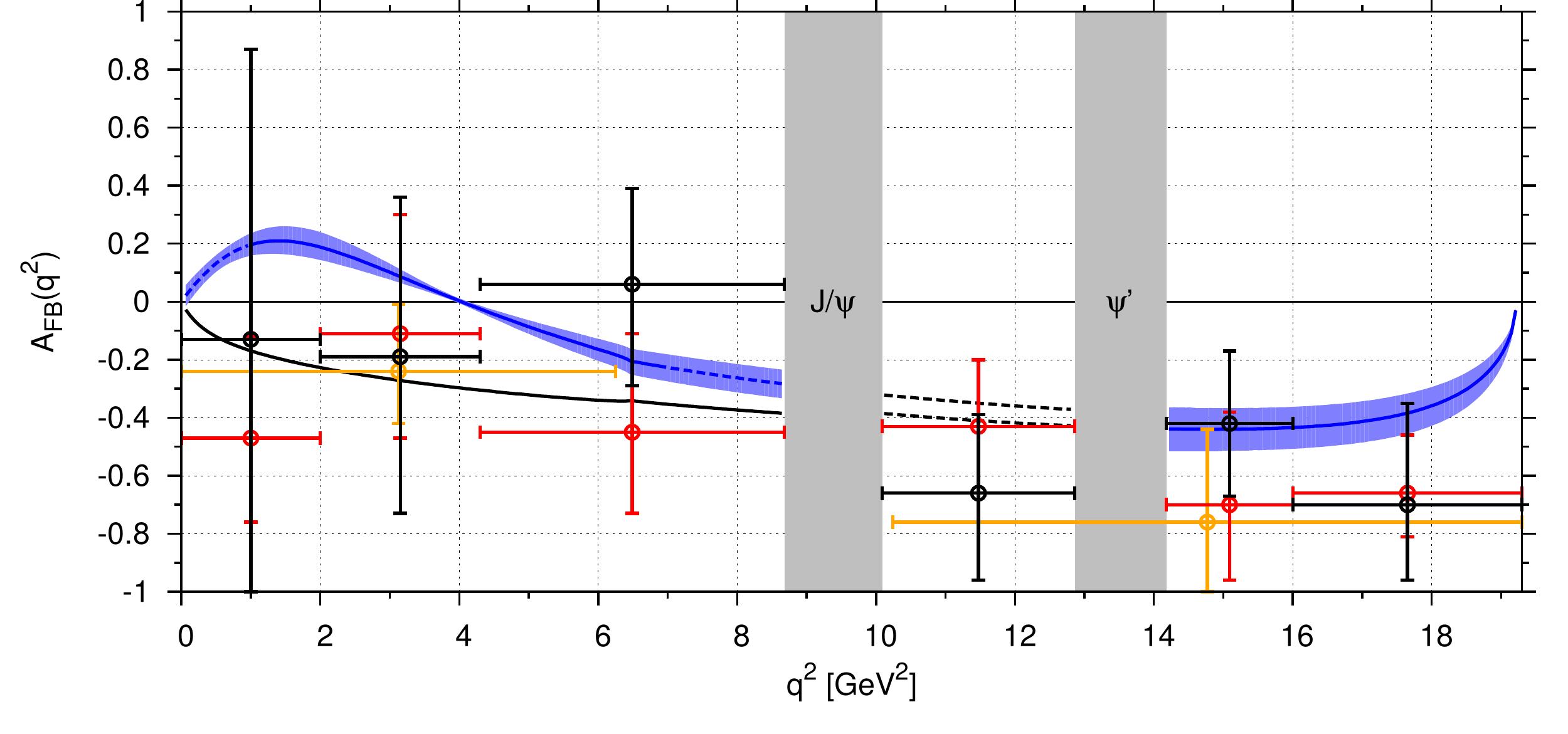,width=6.0cm,height=0.75\linewidth,,angle
=270}} 
\vskip -0.0truein
\end{center}
\caption{$A_{\rm FB}(B \to K^*  \mu^+ \mu^-)$ in the standard model (blue band) versus data:  CDF '10 with $4.4 \mbox{fb}^{-1}$ (black),  BaBar '08 (gold) and  Belle '09 (red). The black
solid curve corresponds to  $C_7 =-C_7^{\rm SM}$. 
 Figure taken  from [6].
 }
 \label{fig:afb}
\end{figure}

Because of the different theory frameworks applicable to the low-$q^2$  and high-$q^2$ region, as well as the  resonance veto, appropriately $q^2$-binned data are vital for  precisely testing the standard model with exclusive semileptonic $b \to s \mu^+ \mu^-$ modes.
The current situation is exemplified in Figure \ref{fig:afb} for the forward-backward asymmetry $A_{\rm FB}$ in $B \to K^*  \mu^+ \mu^-$ decays.

\subsection{The High-$q^2$ Region  \label{subsec:loreco}}

The OPE  \cite{Grinstein:2004vb}  in $1/Q$, $Q=\{m_b, \sqrt{q^2}\}$ is
combined with the improved  heavy quark  form factor  relations \cite{Grinstein:2002cz} 
between the dipole form factors $T_{1,2,3}$ and  vector ones $V,A_{1,2}$.
To leading order in $1/m_b$  and including radiative corrections
\begin{equation} 
  \label{eq:ffrelations}
    T_1(q^2)
        = \kappa V(q^2), ~~~~~
   T_2(q^2)
        = \kappa A_1(q^2), ~~~~~
   T_3(q^2)
         = \kappa A_2(q^2) \frac{m_B^2}{q^2},
\end{equation}
where $\kappa  = 1 - 2 \alpha_s/ 3\pi  \ln \left(\mu/m_b\right) \simeq 1$. The heavy quark-based OPE is  powerful because it predicts
 a simple transversity structure for the $B \to K^*  \mu^+ \mu^-$ decay amplitudes:
Each of the  transversity amplitudes $A_{i}^{L/R}$, $ i=0,\perp,||$ factorizes  into
universal short-distance $C^{L/R}$ and form factor coefficients $f_i$ as \cite{Bobeth:2010wg} 
\footnote{Assuming standard model-type operators commonly termed ${\cal{O}}_7, {\cal{O}}_9$ and ${\cal{O}}_{10}$.}
\begin{equation} \label{eq:fact}
A_{i}^{L/R} \propto C^{L/R} \cdot f_i ,
\end{equation}
up to corrections of order $\alpha_s \Lambda/m_b$ and $(C_7/C_9) \Lambda/m_b$ that is, a few percent.
This in turn allows to design high-$q^2$ observables which are \cite{Bobeth:2010wg}
\cite{Bobeth:2011gi}
\begin{enumerate}
\item independent of the form factors ($H_T^{(2,3)}$, $a_{\rm CP}^{(i)}$); note that 
$H_T^{(3)}$ probes the same  short-distance physics as  $A_{\rm FB}$ while having
a significantly smaller theoretical uncertainty.
\item independent of the short-distance coefficients ($f_i/f_j$) and test  the form factors at low recoil, for instance, the ratio $V/A_1 \propto \sqrt{(2 J_2^s + J_3)/(2 J_2^s - J_3)}$. 
An extraction from data can be used to compare against theory predicitions from lattice
 \cite{Liu:2011ra} or other means.
\item independent of neither short-distance nor form factor coefficients and test the theoretical low recoil framework, such as
\begin{equation} \label{eq:opetest}
H_T^{(1)} =1 , ~~~~~~~ J_7=0.
\end{equation}
\end{enumerate}

The new form factor-free high-$q^2$ observables $H_T^{(i)}$ are defined in terms of the transversity amplitudes as
\begin{eqnarray}
  H_T^{(1)}(q^2) & =  
    \frac{{\rm Re}( A_0^L A_\parallel^{L*} + A_0^{R*} A_\parallel^R)}
     {\sqrt{\big(\vert A_0^L\vert^2 + \vert A_0^R\vert^2\big)
            \big(\vert A_\parallel^L\vert^2 + \vert A_\parallel^R\vert^2\big)}}
  = \frac{\sqrt{2} J_4}{\sqrt{- J_2^c \left(2 J_2^s - J_3\right)}} ,
  \label{eq:def:HT1}
\\
  H_T^{(2)} (q^2)& = 
    \frac{{\rm Re}( A_0^L A_\perp^{L*} - A_0^{R*} A_\perp^R)}
     {\sqrt{\big(\vert A_0^L\vert^2 + \vert A_0^R\vert^2\big)
            \big(\vert A_\perp^L\vert^2 + \vert A_\perp^R\vert^2\big)}}
  = \frac{\beta_l J_5}{\sqrt{-2 J_2^c \left(2 J_2^s + J_3\right)}} ,
  \label{eq:def:HT2}
\\
  H_T^{(3)}(q^2) & =
    \frac{{\rm Re}( A_\parallel^L A_\perp^{L*} - A_\parallel^{R*} A_\perp^R)}
     {\sqrt{\big(\vert A_\parallel^L\vert^2 + \vert A_\parallel^R\vert^2\big)
            \big(\vert A_\perp^L\vert^2 + \vert A_\perp^R\vert^2\big)}}
  = \frac{\beta_l J_6}{2 \sqrt{(2 J_2^s)^2 - J_3^2}} ,
  \label{eq:def:HT3}
\end{eqnarray} 
 or likewise  the angular coefficients $J_i(q^2)$.  Ways to extract the latter  
from single or double differential angular distributions in
$ B \to K^* ( \to K \pi) \mu^+ \mu^-$  decays  have been given in \cite{Bobeth:2008ij}.

Eq.~(\ref{eq:fact}) further limits the number of independent
CP-asymmetries  in $B \to K^* \mu^+ \mu^-$ decays to three, one related to the rate, $a_{\rm CP}^{(1)}$,  one related to $A_{\rm FB}$, $a_{\rm CP}^{(2)}$, or  with a more favorable normalization related to $H_T^{(2,3)}$, $a_{\rm CP}^{(3)}$, and one from meson mixing, $a_{\rm CP}^{mix}$ \cite{Bobeth:2011gi}.
 
 Beylich {\it et al.}~\cite{Beylich:2011aq} recently  proposed a local expansion without
 engaging heavy quark effective theory. In their OPE, Eq.~(\ref{eq:fact}) is not manifest, hence the aforementioned symmetry-based  high-$q^2$  predictions 1.~-- 3. are no longer explicit, however,
 the OPE itself has a simpler structure.
 It will become most useful once all $B \to K^*$ form factors are known with sufficient accuracy.
 
The treatment of the high-$q^2$ region is based on an OPE, whose
performance can be tested by checking {\it e.g.}, Eq.~(\ref{eq:opetest}).
The OPE is supported by consistency between the constraints obtained from excluding
and using only the high-$q^2$ region data \cite{Bobeth:2010wg}, and by a recent model-study~\cite{Beylich:2011aq}.

\subsection{Angular Distributions \label{subsec:anganal}}

With high event rates at the horizon the angular analysis
with an on-shell decaying $K^* \to K \pi$ \cite{Kruger:1999xa} has 
received recently a lot of  attention as a tool to
maximize the extraction of  physics   from $B \to K^* \mu^+ \mu^-$ decays 
\cite{Bobeth:2010wg,Bobeth:2011gi,Bobeth:2008ij,Egede:2008uy,Altmannshofer:2008dz,Egede:2010zc}.
In a full angular analysis  the quartic
 differential decay distribution $d^4 \Gamma$ factorizes into $q^2$-dependent angular coefficients $J_i$ and trigonometric functions of
  three angles:
 $\thl$, the angle between the $l^-$  and the $\bar B$ in the dilepton CMS,
$\thK$, the angle between the $K$ and the $\bar B$ in the $K^*$-CMS and 
$\phi$, the angle between the normals of the $K \pi$ and $l^+ l^-$ plane 
\begin{equation} d^4\Gamma =\frac{3}{8 \pi} J(q^2, \thl, \thK, \phi) dq^2 d
 \cos \thl d \cos \thK d\phi ,
 \end{equation}
  where
 \begin{eqnarray}
  J(q^2, \thl, \thK, \phi)& = &J_1^s \sin^2\thK + J_1^c \cos^2\thK
      + (J_2^s \sin^2\thK + J_2^c \cos^2\thK) \cos 2\thl
\nonumber \\       
    & +& J_3 \sin^2\thK \sin^2\thl \cos 2\phi 
      + J_4 \sin 2\thK \sin 2\thl \cos\phi 
      + J_5 \sin 2\thK \sin\thl \cos\phi
\nonumber \\      
    & + &J_6 \sin^2\thK \cos\thl 
      + J_7 \sin 2\thK \sin\thl \sin\phi
\nonumber \\ 
    & + &J_8 \sin 2\thK \sin 2\thl \sin\phi
      + J_9 \sin^2\thK \sin^2\thl \sin 2\phi , ~~~~~~J_i =J_i(q^2) .
  \label{eq:I:func}
\end{eqnarray}

The angular analysis offers opportunities  for searches for  beyond the standard model (BSM) CP-violation. The angular distribution $d^4 \bar \Gamma$ of the
CP-conjugate decays is obtained from $d^4 \Gamma$ after flipping the sign of the CP phases and by replacing $J_{1,2,3,4,7} \to \bar J_{1,2,3,4,7}$ and
$J_{5,6,8,9} \to - \bar J_{5,6,8,9}$.
Several CP-asymmetries $A_i \propto J_i-\bar J_i$ can be obtained.
Highlights include  \cite{Bobeth:2011gi,Bobeth:2008ij}:
$A_{3,9}$ vanish in the standard model by helicity conservation, hence, they are
sensitive to right-handed currents.
$A_{3,9,(6)}$ can be extracted from a single-differential distribution in $\phi (\thl)$.
$A_{7,8,9}$ are (naive) $T$-odd and not suppressed by small strong phases; they can be
 order one with order one  BSM CP-phases.
$A_{5,6,8,9}$ and $a_{\rm CP}^{(3)}$ are CP-odd and can be extracted  without tagging from $\Gamma +\bar \Gamma$;
this is advantageous for $B_s, \bar B_s \to  (\Phi \to K^+ K^-) \mu^+ \mu^-$ decays which are not
self-tagging; time-integrated measurements are possible as well.
Note that in the standard model  all $b \to s$ CP-asymmetries  are doubly Cabibbo-suppressed
and small.
At high $q^2$, due to  Eq.~(\ref{eq:fact}),  $A_{7,8,9}=0$. The angular distribution in $B \to K \mu^+ \mu^-$ decays  involves only one angle $\thl$ and is simpler  \cite{Bobeth:2007dw}.

\section{BSM Implications \label{sec:bsm}}

\begin{figure}
\begin{center}
\vskip -0.0truein
{\epsfig{file=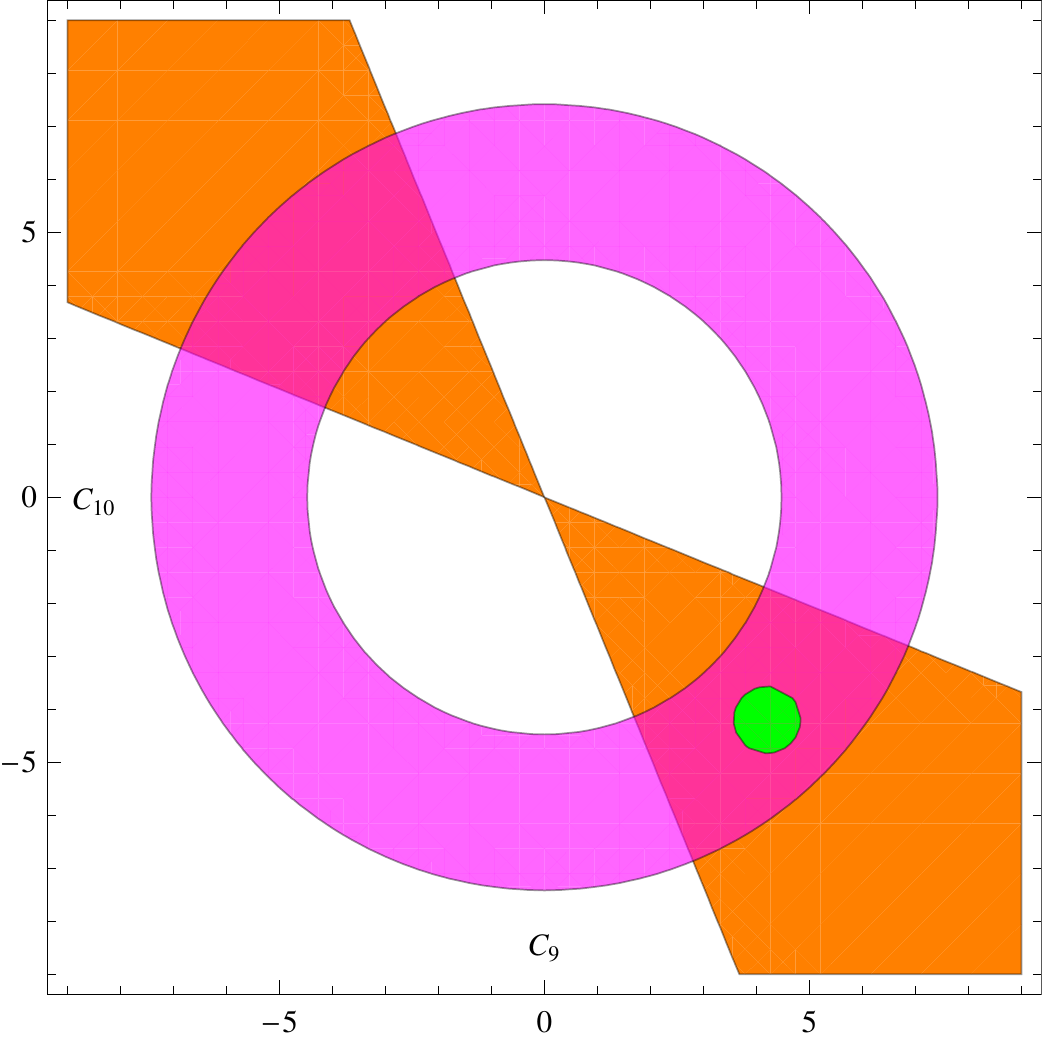,width=6.5cm,height=0.39\linewidth,,angle=0}}
\hspace{1.0cm}
{\epsfig{file=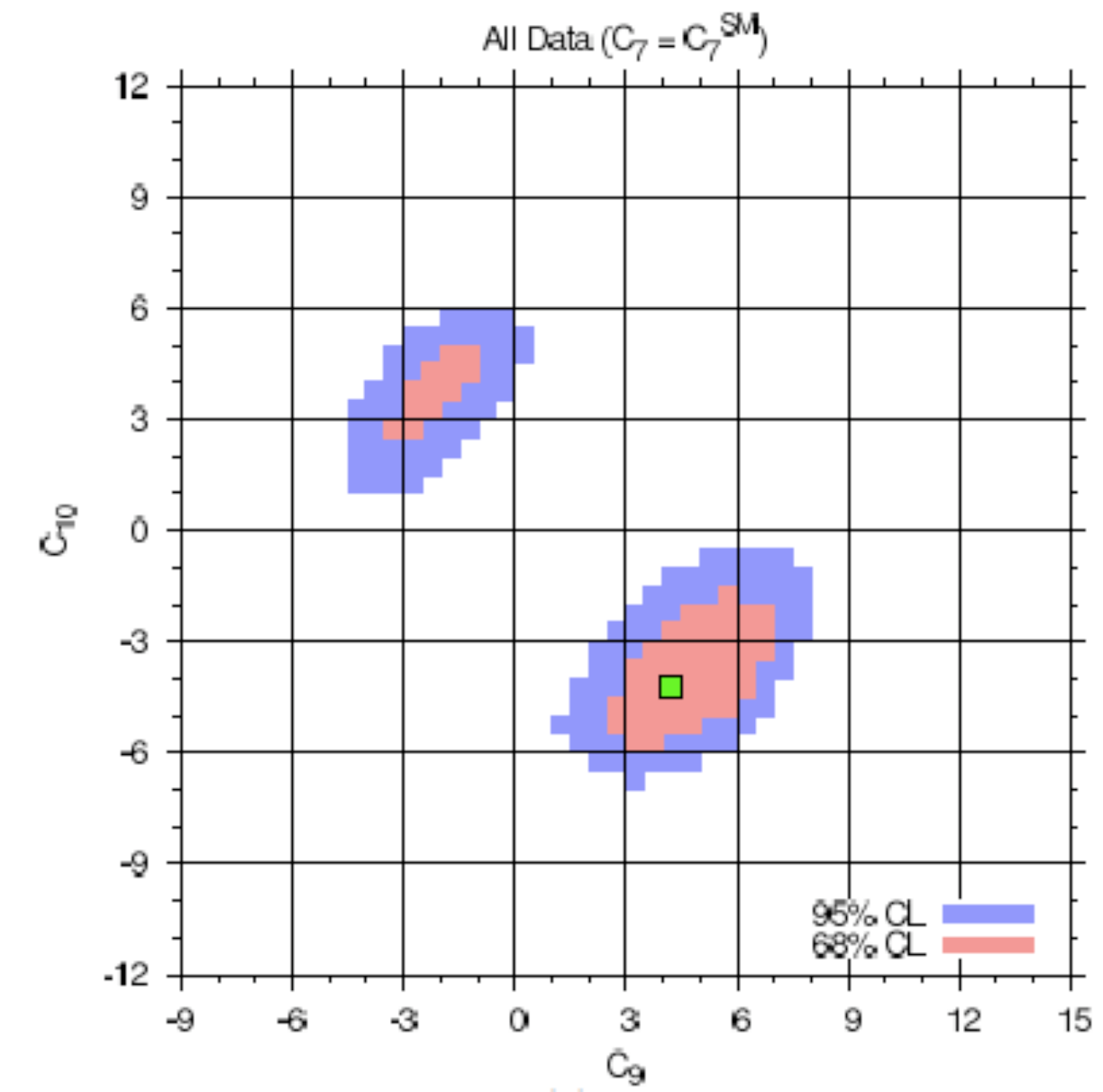,width=6.9cm,height=0.39\linewidth,,angle=0}}  
\end{center}
\caption{
Todays model-independent bounds on real-valued $C_9,C_{10}$ from $b \to s \mu^+ \mu^-$ decays for $C_7 \simeq C_7^{\rm SM} <0$.  The left-handed plot is schematic only and
illustrates the allowed regions from branching ratio measurements (magenta ring) and
$A_{\rm FB}$ determinations at large $q^2$ (orange wedges).
 The green dot  corresponds to  $(C_9^{\rm SM},C_{10}^{\rm SM})$. 
The right-handed plot with the allowed 68 and 95 \% C.L. regions is 
taken from Ref.
[6].
There are similar plots for $C_7\simeq -C_7^{\rm SM} >0$, hence in total four disconnected
allowed regions in $C_7,C_9$ and $C_{10}$.}
\vskip -0.0truein
\label{fig:mia}
\end{figure}

Measurements of $B \to K^{*} \mu^+ \mu^-$  observables
place model-independent constraints on the Wilson coefficients of the four-fermi operators
${\cal{O}}_{9,10}$.  Assuming real-valued coefficients $C_9,C_{10}$
the constraints from branching ratio and $A_{\rm FB}$ spectra at high $q^2$ are illustrated 
 in the left-handed plot of Figure \ref{fig:mia}. 
The high-$q^2$ constraints from  $A_{\rm FB}\sim Re(C_{10} C_9^*)$ are orthogonal to the ones from the branching ratio $\sim |C_9|^2 +|C_{10}|^2$.
 The magnitude of $C_7$ is fixed by the measured $B \to X_s \gamma$ branching ratio
to be near its standard model value.
  
 The outcome of a recent analysis using $B \to K^{*} \mu^+ \mu^-$   data \cite{Aubert:2008ps,:2009zv,Aaltonen:2011cn} as well as the constraints from $B \to X_s l^+l^-$ decays 
is shown in the right-handed plot of Figure \ref{fig:mia} \cite{Bobeth:2010wg}.
The constraints from  $A_{\rm FB}$ at high $q^2$ significantly improve the scan. They are manifest in the plane by selecting arcs from the area allowed by the various branching ratio measurements.
Because of the  $Re(C_7 C_9^*)$ interference term in
${\cal{B}}(B \to X_s l^+l^-)$ the ambiguity in the disconnected allowed regions is mildly broken.
The allowed regions include the
standard model, but order one deviations in all three Wilson coefficients $C_7,C_9$ and $C_{10}$ are allowed as well.

The experimental situation of $A_{\rm FB}$ at low $q^2$, unlike the one at high $q^2$, 
is currently not settled, see Figure \ref{fig:afb}. To find out whether there is a zero-crossing of the 
$A_{\rm FB}$  at low $q^2$ as predicted by the standard model $q_0^2|_{\rm SM}  =4.36^{+0.33}_{-0.31} \,  {\mbox{GeV}}^2$ (for $B^0 \to K^{*0} \mu^+ \mu^-$ ),
$q_0^2|_{\rm SM}  =4.15 \pm 0.27 \,  {\mbox{GeV}}^2$ (for $B^+ \to K^{*+} \mu^+ \mu^-$ ) \cite{Beneke:2004dp} 
and likewise for $B \to X_s l^+l^-$ decays
$q_0^2|_{\rm SM}  =(3.34 \ldots 3.40)^{+0.22}_{-0.25}  \,  {\mbox{GeV}}^2$ \cite{Bell:2010mg} 
remains  a central goal for  next years $b$-physics programs.

\begin{figure}
\begin{center}
\vskip -0.0truein
{\epsfig{file=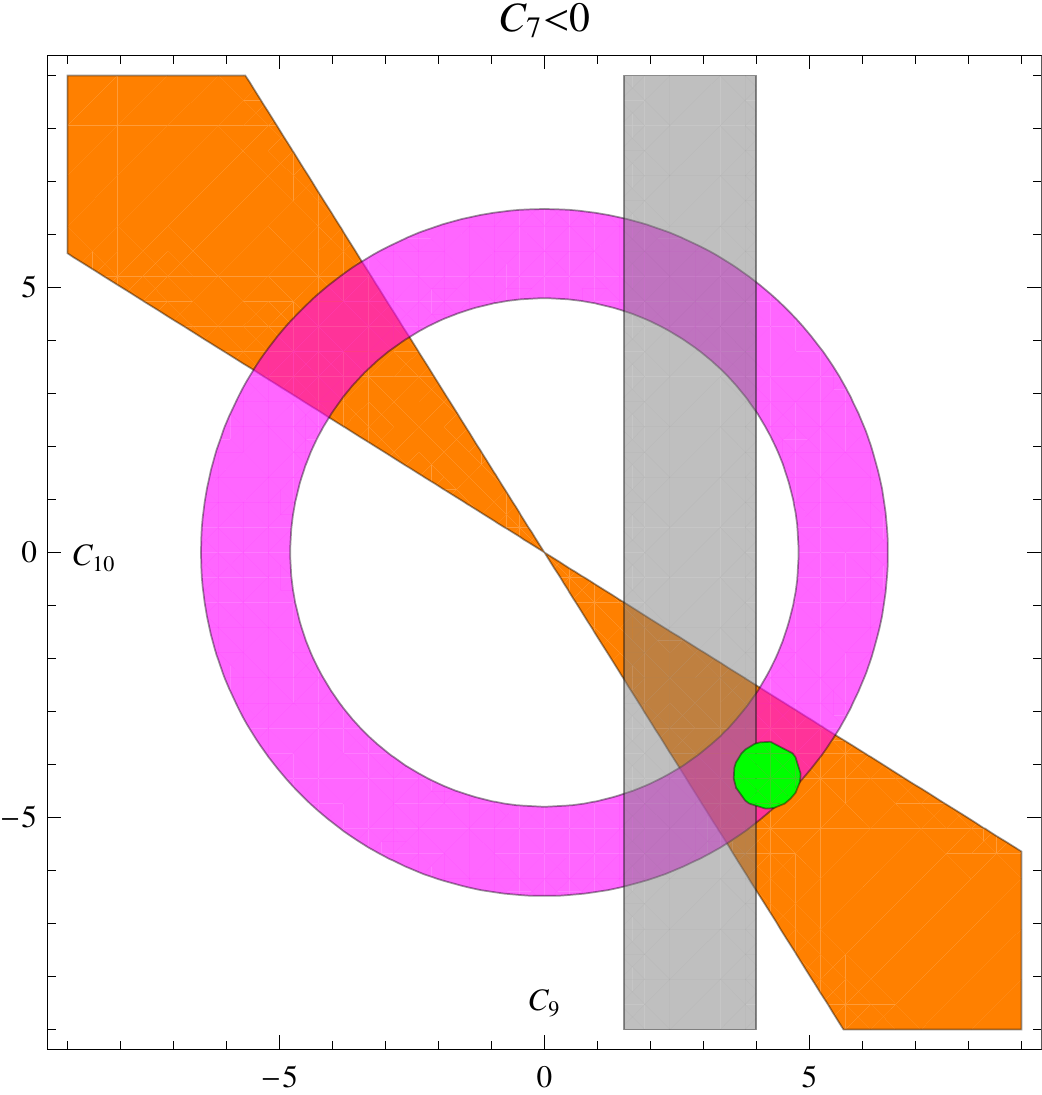,width=6.5cm,height=0.39\linewidth,,angle=0}} 
\hspace{1.0cm}
{\epsfig{file=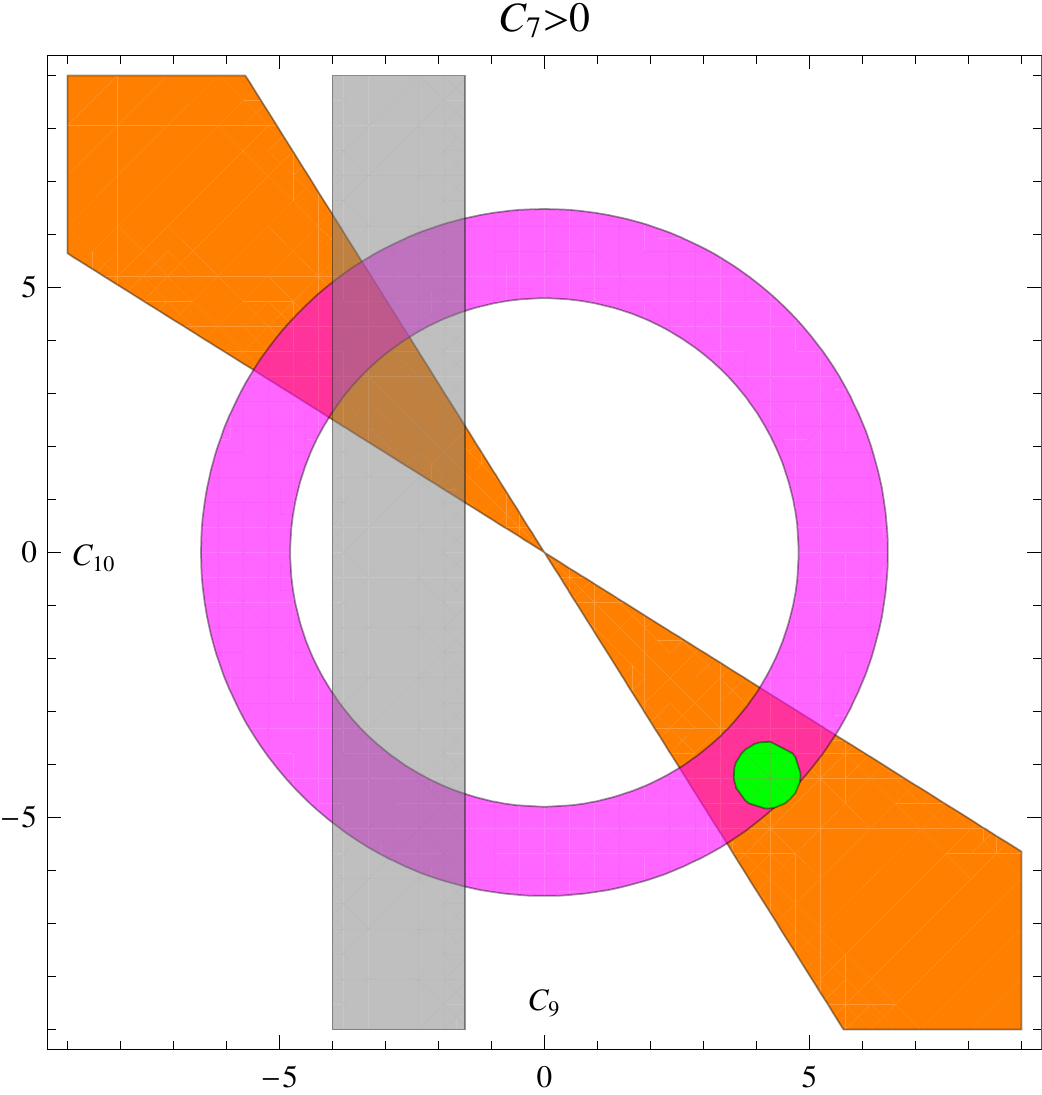,width=6.5cm,height=0.39\linewidth,,angle=0}} 
\end{center}
\caption{ Future 
scenario of the model-independent bounds on real-valued $C_9,C_{10}$ from 
$b \to s \mu^+ \mu^-$ decays for $C_7\simeq C_7^{\rm SM} <0$ (left-handed plot) and $C_7\simeq - C_7^{\rm SM} >0$ (right-handed plot).
The grey vertical bands denote the constraints arising if an
$A_{\rm FB}$ zero at low $q^2$ could be established. There remain 
two allowed disconnected regions.}
\vskip -0.0truein
\label{fig:mia2}
\end{figure}

A future analysis assuming the  existence and determination of the $A_{\rm FB}$ zero at low $q^2$  is given in Figure \ref{fig:mia2}. The establishment of the zero reduces the four-fold 
ambiguity to two.
Resolving the last ambiguity requires precision studies sensitive to the contributions from
four-quark operators which are commonly absorbed in the effective coefficients $C_i^{\rm eff}$.
Assuming vanishing or very small CP phases a lower bound on the position of the 
$A_{\rm FB}(B \to K^{*} \mu^+ \mu^-)$  zero
can be derived from the respective upper bounds on $|C_9|$  \cite{Bobeth:2010wg}. Very roughly, $q_0^2 \simeq q_0^2|_{\rm SM} |C_9^{\rm SM}|/|C_9^{\rm max}|
 \gsim 2\, {\mbox{GeV}}^2$.

Model-independent $\Delta B=1$ BSM implications can be drawn using an effective theory\footnote{Thanks to Gilad Perez for suggesting this.}
\begin{eqnarray} 
{\cal{H}}_{\rm eff}= 
\sum_{i} \frac{\tilde c _i}{\Lambda_{\rm NP}^2}  \widetilde O_i  \,,~~~~
\widetilde O_{10}=  \bar s  \gamma_\mu (1-\gamma_5)b \, \bar \mu \gamma^\mu \gamma_5 \mu \, .
\end{eqnarray}
Assuming new physics at the scale $\Lambda_{\rm NP}=1$ TeV
the coefficient of the higher dimensional operator $\tilde O_{10}$ needs a (flavor) suppression 
as strong as $|\tilde c_{10}| <  2 \cdot 10^{-3} \, (5 \cdot 10^{-3})$.
If one assumes no suppression at all, $|\tilde c_{10}| =1 $, the scale of new physics is pushed up to
\begin{equation}
\Lambda_{\rm NP} > 26 \, \mbox{TeV} ~(15 \, \mbox{TeV}) \, .
\end{equation}
The bounds are obtained  at 95 \% C.L. \cite{Bobeth:2010wg}.
The first numbers correspond to ${\rm max} |C_{10} -C_{10}^{\rm SM}|$ from the
nearby solution, that is, from the allowed region including the standard model 
whereas the weaker bounds in parentheses  are obtained from the far away region, not connected 
to $(C_9^{\rm SM},C_{10}^{\rm SM})$.

A recent complex-valued scan  \cite{eos} in $C_{7,9,10}$  returns the allowed  68\% C.L. (95\% C.L.) ranges \cite{Bobeth:2011gi}
\begin{eqnarray} \nonumber
  0.8 \leq  |C_{9}| \leq 6.8 & (0.0 \leq  |C_9| \leq 7.8) \, ,\\
  1.8 \leq  |C_{10}| \leq 5.5 & (0.8 \leq  |C_{10}| \leq 6.3)  \, .
  \label{eq:c10limits}
  \end{eqnarray}
with some of the  lower bounds being sensitive to the discretization 
of the scan.
 The  constraints on the CP phases are not very strong,
approximately
$ \frac{\pi}{2} \lsim \,{\rm arg}\,(C_{9} C^*_{10}) \lsim \frac{3\pi}{2} $
at 68\% C.L. \cite{Bobeth:2011gi}.
  
  Eqs.~(\ref{eq:c10limits}) imply a maximal enhancement of the  $\bar{B}_s \to \mu^+\mu^-$ 
branching ratio ${\cal{B}}(\bar{B}_s \to \mu^+\mu^-) \propto f_{B_s}^2 |C_{10}|^2$  with respect to its standard model value by a factor 2.3. The dominant uncertainty of the standard model
prediction is stemming from the  decay constant of the $B_s$ meson.
  Using  \cite{Gamiz:2009ku} \cite{Simone:2010zz} gives 
  \cite{Bobeth:2011gi}
\begin{eqnarray} \nonumber
  {\cal{B}}(\bar{B}_s \to \mu^+\mu^-)_{\rm SM} =(3.1 \pm 0.6) \times 10^{-9} ,& 
  f_{Bs}=231(15)(4)\,  \mbox{MeV} ~~(\mbox{Gamiz et al '09}) ,
\\
{\cal{B}}(\bar{B}_s \to \mu^+\mu^-)_{\rm SM} =(3.8 \pm 0.4) \times 10^{-9} ,&
f_{Bs}=256(6)(6)\, \mbox{MeV}  ~~(\mbox{Simone et al '10}) . \label{eq:brsm}
    \end{eqnarray}
It follows the upper limit  
(at 95 \% C.L.)  \cite{Bobeth:2011gi}
\begin{equation}
{\cal{B}}(\bar{B}_s \to \mu^+\mu^-) <10 \times 10^{-9} \, ,
\end{equation}
which could be invalidated by sizable contributions from scalar and pseudo-scalar operators not considered here. Experimentally, ${\cal{B}}(\bar{B}_s \to \mu^+\mu^-) <43 \times 10^{-9}$ at 95 \% C.L.  from CDF~\cite{cdf}.

\section{Outlook}

At this stage first results for basic decay distributions and asymmetries of exclusive $b \to s l^+l^-$ modes have become available.
With more data soon and shrinking error bars the constraints from the $\Delta B=1$ 
analysis will tighten. Steep progress in the BSM reach is expected from additional and complementary observables which could remove -- or verify --  currently allowed solutions far away from the standard model. 
A useful nearer term observable in this regard  is $A_{\rm FB}$ at low recoil, perhaps also combined with  improved constraints from  the $B \to X_s l^+l^-$ branching ratio. 
Further observables designed with good theory properties are accessible by angular analysis, which is promising for higher statistics searches.

BSM models often induce  operators beyond those of the standard model.
These include right-handed currents, enhanced
scalar and pseudo-scalar couplings or lepton-flavor non-universal effects, and can {\it e.g.}~be searched for with, respectively, transverse asymmetries \cite{Egede:2008uy}, $\bar B_s \to \mu^+ \mu^-$ or by comparing $l=e$ to $\mu$ modes \cite{Bobeth:2007dw}.
${\cal{O}}(1)$ BSM CP phases can show up as ${\cal{O}}(1)$ $T$-odd CP asymmetries $A_{7,8,9}$
\cite{Bobeth:2008ij}. Angular analysis becomes most powerful and essential here.

Dimuons provide great opportunities for
LHC(b) and the Tevatron. They could also be studied at future super flavor $e^+ e^-$ factories, which moreover have good capabilities to investigate dielectron and inclusive modes, and missing energy searches covering $l=\nu$ or possibly $l=\tau$.

\section*{Acknowledgments}
GH is happy to thank her collaborators from the EOS team and the organizers of the 2011 Moriond Electroweak session for the invitation to the meeting.
The works presented here are supported in part by the Bundesministerium f\"ur Bildung und
Forschung (BMBF).

\end{document}